%% file: fabric.tex
\documentclass[sigconf]{acmart}

\renewcommand\footnotetextcopyrightpermission[1]{} 
\usepackage{enumitem}
\setitemize{noitemsep,parsep=0pt,partopsep=0pt}

\usepackage{xspace}
\PassOptionsToPackage{hyphens}{url}\usepackage{hyperref}
\usepackage{graphicx}
\usepackage{footmisc}
\usepackage{caption, subcaption}
\usepackage{makecell}

\newcommand{\HLF}{Fabric\xspace}
\newcommand{\coin}{Fabcoin\xspace}

\newcommand{\var}[1]{\textit{#1}}
\newcommand{\op}[1]{\textit{#1}}

\providecommand{\nondet}{non-de\-ter\-min\-istic\xspace}
\begin{document}

\title{Hyperledger Fabric: A Distributed Operating System for Permissioned Blockchains }

\author{Elli Androulaki}
\author{Artem Barger}
\author{Vita Bortnikov}
\affiliation{IBM}  
\author{Christian Cachin}
\author{Konstantinos Christidis}
\author{Angelo De Caro} 
\author{David Enyeart}
\affiliation{IBM} 
\author{Christopher Ferris} 
\author{Gennady Laventman} 
\author{Yacov Manevich} 
\affiliation{IBM} 
\author{Srinivasan Muralidharan}
\affiliation{State Street Corp.}
\authornote{Work done at IBM.}
\author{Chet Murthy}
\affiliation{\mbox{}}
\authornotemark[1]
\author{Binh Nguyen} 
\affiliation{State Street Corp.}
\authornotemark[1]
\author{Manish Sethi} 
\author{Gari Singh} 
\author{Keith Smith}
\author{Alessandro Sorniotti}
\affiliation{IBM} 
\author{Chrysoula Stathakopoulou} 
\author{Marko Vukoli\'c}
\author{Sharon Weed Cocco}
\author{Jason Yellick}
\thanks{\copyright\ Copyright held by the owner/author(s).}
\affiliation{IBM\vspace*{1.8cm}}


\begin{abstract}

	Fabric is a modular and extensible open-source system for deploying and operating permissioned blockchains and one of the Hyperledger projects hosted by the Linux Foundation (\url{www.hyperledger.org}).

	\HLF is the first truly extensible blockchain system for running distributed applications.  It supports modular consensus protocols, which allows the system to be tailored to particular use cases and trust models. \HLF is also the first blockchain system that runs distributed applications written in standard, general-purpose programming languages, without systemic dependency on a native cryptocurrency.  This stands in sharp contrast to existing blockchain platforms that require ``smart-contracts'' to be written in domain-specific languages or rely on a cryptocurrency.  \HLF realizes the permissioned model using a portable notion of membership, which may be integrated with industry-standard identity management.  To support such flexibility, \HLF introduces an entirely novel blockchain design and revamps the way blockchains cope with non-determinism, resource exhaustion, and performance attacks.

	This paper describes \HLF, its architecture, the rationale behind various design decisions,
	its most prominent implementation aspects, as well as its distributed application programming model. We further evaluate \HLF by implementing and benchmarking a Bitcoin-inspired digital currency. We show that \HLF achieves end-to-end throughput of more than 3500 transactions per second in certain popular deployment configurations, with sub-second latency, scaling well to over 100 peers.

\end{abstract}

\maketitle
\renewcommand{\shortauthors}{E. Androulaki et al.}
\thispagestyle{empty}

\input{introduction}

\input{background}

\input{architecture}
\input{components}

\input{evaluation}

\input{apps}

\input{relatedwork}

\input{conclusion}

\newpage

\bibliographystyle{abbrv}
\bibliography{fault_tolerance,paxos,blockchain,gossip,ordering,ledger}

\end{document}

%% file: introduction.tex
\section{Introduction}
\label{sec:introduction}

A blockchain can be defined as an immutable \emph{ledger} for recording \emph{transactions}, maintained within a distributed network of mutually untrusting \emph{peers}. Every peer maintains a copy of the ledger. The peers execute a \emph{consensus protocol} to validate transactions, group them into blocks, and build a hash chain over the blocks. This process forms the ledger by ordering the transactions, as is necessary for consistency. Blockchains have emerged with Bitcoin \cite{BitcoinURL}
and are widely regarded as a promising technology to run trusted exchanges in the digital world.

In a \emph{public} or \emph{permissionless} blockchain anyone can participate without a specific identity. Public blockchains typically involve a native cryptocurrency and often use consensus based on ``proof of work'' (PoW) and economic incentives. \emph{Permissioned} blockchains, on the other hand, run a blockchain among a set of known, identified participants. A permissioned blockchain provides a way to secure the interactions among a group of entities that have a common goal but which do not fully trust each other, such as businesses that exchange funds, goods, or information. By relying on the identities of the peers, a permissioned blockchain can use traditional Byzantine-fault tolerant (BFT) consensus.

Blockchains may execute arbitrary, programmable transaction logic in the form of \emph{smart contracts}, as exemplified by Ethereum \cite{EthereumURL}.
The scripts in Bitcoin were a predecessor of the concept. A smart contract\hfill functions\hfill as\hfill a\hfill \emph{trusted\hfill distributed\hfill application}\hfill and\hfill gains\break
its\hfill security\hfill from\hfill the\hfill blockchain\hfill and\hfill the\hfill underlying\hfill consensus\break
\newpage

\noindent among the peers. This closely resembles the well-known approach of building resilient applications with state-machine replication (SMR)~\cite{Schneider90}. However, blockchains depart from traditional SMR with Byzantine faults in important ways: (1) not only one, but many distributed applications run concurrently; (2) applications may be deployed dynamically and by anyone; and (3) the application code is untrusted, potentially even malicious. These differences necessitate new designs.

Many existing smart-contract blockchains follow the blueprint of SMR~\cite{Schneider90} and implement so-called \emph{active} replication~\cite{CharronBostPS10}: a protocol for \emph{consensus} or \emph{atomic broadcast} first orders the transactions and propagates them to all peers; and second, each peer executes the transactions sequentially. We call this the \emph{order-execute architecture}; it requires all peers to execute every transaction and all transactions to be deterministic. The order-execute architecture can be found in virtually all existing blockchain systems, ranging from public ones such as Ethereum (with PoW-based consensus) to permissioned ones (with BFT-type consensus) such as Tendermint \cite{TendermintURL},
Chain \cite{ChainURL}
and Quorum \cite{QuorumURL}.
Although the order-execute design is not immediately apparent in all systems, because the additional transaction validation step may blur it, its limitations are inherent in all: every peer executes every transaction and transactions must be deterministic.

Prior permissioned blockchains suffer from many limitations, which often stem from their permissionless relatives or from using the order-execute architecture.  In particular:
\begin{itemize}
\item Consensus is \emph{hard-coded} within the platform, which contradicts the well-established understanding that there is no ``one-size-fits-all'' (BFT) consensus protocol~\cite{SinghDMDR08};
\item The \emph{trust model} of transaction validation is determined by the consensus protocol and cannot be adapted to the requirements of the smart contract;
\item Smart contracts must be written in a \emph{fixed, non-standard, or domain-specific language}, which hinders wide-spread adoption and may lead to programming errors;
\item The \emph{sequential execution} of all transactions by all peers \emph{limits performance}, and complex measures are needed to prevent denial-of-service attacks against the platform originating from untrusted contracts (such as accounting for runtime with ``gas'' in Ethereum);
\item Transactions must be \emph{deterministic}, which can be difficult to ensure programmatically;
\item Every smart contract runs on \emph{all} peers, which is at odds with \emph{confidentiality}, and prohibits the dissemination of contract code and state to a subset of peers.
\end{itemize}

In this paper we describe \emph{Hyperledger Fabric} or simply \emph{Fabric}, an open-source \cite{FabricCodeURL}
blockchain platform that overcomes these limitations.  \HLF is one of the projects of Hyperledger \cite{HyperledgerURL}
under the auspices of the Linux Foundation \cite{LinuxFoundationURL}.
\HLF is used in more than 400 prototypes, proofs-of-concept, and in production distributed-ledger systems, across different industries and use cases.  These use cases include but are not limited to areas such as dispute resolution, trade logistics, FX netting, food safety, contract management, diamond provenance, rewards point management, low liquidity securities trading and settlement, identity management, and settlement through digital currency. 

Fabric introduces a new blockchain architecture aiming at resiliency, flexibility, scalability, and confidentiality.  Designed as a modular and extensible general-purpose permissioned blockchain, Fabric is the first blockchain system to support the execution of distributed applications written in \emph{standard programming languages}, in a way that allows them to be executed consistently across many nodes, giving impression of execution on a single globally-distributed blockchain computer.  This makes Fabric the first \emph{distributed operating system} \cite{Tanenbaum93} for permissioned blockchains.

The architecture of Fabric follows a novel \emph{execute-order-validate} paradigm for distributed execution of untrusted code in an untrusted environment.  It separates the transaction flow into three steps, which may be run on different entities in the system: (1) \emph{executing} a transaction and checking its correctness, thereby \emph{endorsing} it (corresponding to ``transaction validation'' in other blockchains); (2) \emph{ordering} through a consensus protocol, irrespective of transaction semantics; and (3) transaction \emph{validation} per application-specific trust assumptions, which also prevents race conditions due to concurrency.  

This design departs radically from the order-execute paradigm in that Fabric typically executes  transactions before reaching final agreement on their order.  It combines the two well-known approaches to replication, \emph{passive} and \emph{active}, as follows.

First, Fabric uses \emph{passive} or \emph{primary-backup replication}~\cite{Budhiraja:1993:PA,CharronBostPS10} as often found in distributed databases, but with middleware-based asymmetric update processing~\cite{KemmeA00,DBLP:series/synthesis/2010Kemme} and ported to untrusted environments with Byzantine faults.  In Fabric, every transaction is executed (endorsed) only by a subset of the peers, which allows for parallel execution and addresses potential non-determinism, drawing on ``execute-verify'' BFT replication~\cite{Kapritsos0QCAD12}.  A flexible endorsement policy specifies which peers, or how many of them, need to vouch for the correct execution of a given smart contract. 

Second, Fabric incorporates \emph{active replication} in the sense that the transaction's effects on the ledger state are only written after reaching consensus on a total order among them, in the deterministic validation step executed by each peer individually.  This allows Fabric to respect application-specific trust assumptions according to the transaction endorsement.  Moreover, the ordering of state updates is delegated to a modular component for consensus (i.e., atomic broadcast), which is stateless and logically decoupled from the peers that execute transactions and maintain the ledger.  Since consensus is modular, its implementation can be tailored to the trust assumption of a particular deployment.  Although it is readily possible to use the blockchain peers also for implementing consensus, the separation of the two roles adds flexibility and allows one to rely on well-established toolkits for CFT (crash fault-tolerant) or BFT ordering.

Overall, this \emph{hybrid replication} design, which mixes passive and active replication in the Byzantine model, and the \emph{execute-order-validate} paradigm, represent the main innovation in \HLF architecture. They resolve the issues mentioned before and make Fabric a scalable system for permissioned blockchains supporting flexible trust assumptions.

To implement this architecture,  Fabric contains modular building blocks for each of the following components:
\begin{itemize}
\item An \emph{ordering service} atomically broadcasts state updates to peers and establishes consensus on the order of transactions.

\item A \emph{membership service provider} is responsible for associating peers with cryptographic identities. It maintains the permissioned nature of Fabric.
\item An optional \emph{peer-to-peer gossip service} disseminates the blocks output by ordering service  to all peers.
\item \emph{Smart contracts} in \HLF run within a container environment for isolation.  They can be written in standard programming languages but do not have direct access to the ledger state.
\item Each peer locally maintains the \emph{ledger} in the form of the append-only blockchain and as a snapshot of the most recent state in a key-value store.
\end{itemize}

The remainder of this paper describes the architecture of Fabric and our experience with it.  Section~\ref{sec:background} summarizes the state of the art and explains the rationale behind various design decisions.  Section~\ref{sec:architecture} introduces the architecture and the execute-order-validate approach of Fabric in detail, illustrating the transaction execution flow.  
In Section~\ref{sec:components}, the key components of Fabric are defined, in particular, the ordering service, membership service, peer-to-peer gossip, ledger database, and smart-contract API.  Results and insights gained in a performance evaluation of Fabric with a Bitcoin-inspired cryptocurrency, deployed in a cluster and WAN environments on commodity public cloud VMs, are given in Section~\ref{sec:evaluation}.  They show that \HLF achieves, in popular deployment configurations,  throughput of more than 3500 tps, achieving finality \cite{Vukolic15} with latency of a few hundred ms and scaling well to over 100 peers. In Section~\ref{sec:apps} we discuss a few real production use cases of \HLF. 
Finally, Section~\ref{sec:relatedwork} discusses related work.

%% file: background.tex
\section{Background}
\label{sec:background}

\subsection{Order-Execute Architecture for Blockchains}
\label{sec:order-execute}

All previous blockchain systems, permissioned or not, follow the order-execute architecture. This means that the blockchain network orders transactions first, using a consensus protocol, and then executes them in the same order on all peers sequentially.\footnote{In many blockchains with a hard-coded primary application, such as Bitcoin, this transaction execution is called ``transaction validation.'' Here we call this step \emph{transaction execution} to harmonize the terminology.}

\begin{figure}
  \centering
  \includegraphics[width=0.762\columnwidth]{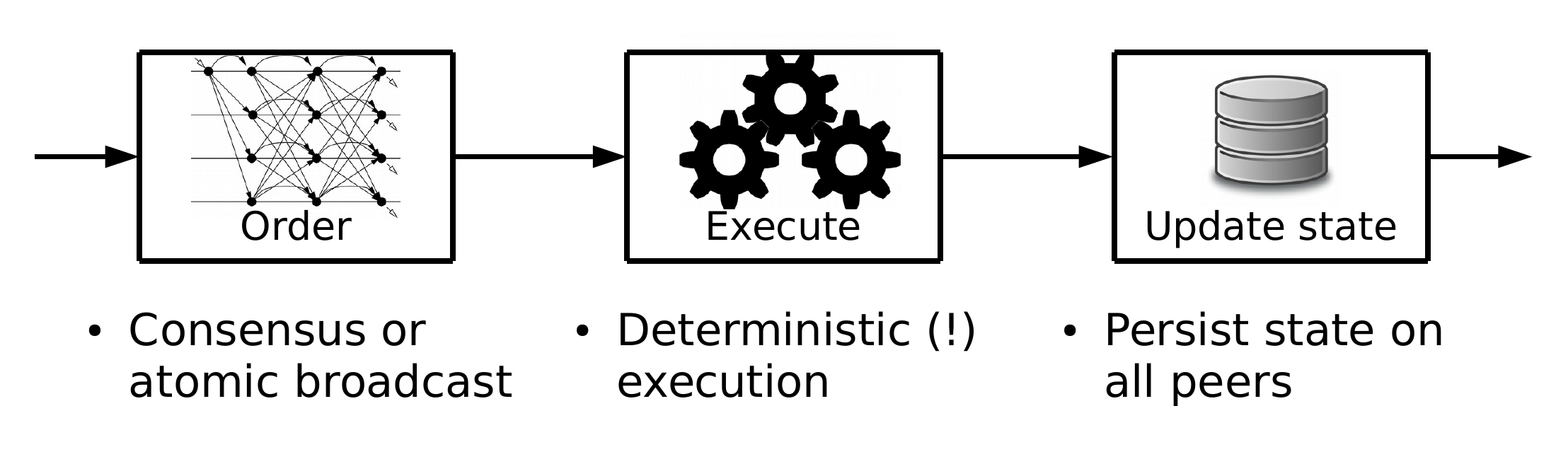}

  \caption{Order-execute architecture in replicated services.}
  \label{fig:order-exec}
\end{figure}

For instance, a PoW-based permissionless blockchain such as Ethereum combines consensus and execution of  transactions as follows: (1) every peer (i.e., a node that participates in consensus)  assembles a block containing valid transactions (to establish validity, this peer already pre-executes those transactions); (2) the peer tries to solve a PoW puzzle~\cite{Nakamoto:Bitcoin}; (3) if the peer is lucky and solves the puzzle, it disseminates the block to the network via a gossip protocol; and (4) every peer receiving the block validates the solution to the puzzle \emph{and} all transactions in the block.  Effectively, every  peer thereby repeats the execution of the lucky peer from its first step.  Moreover, all peers execute the transactions  \emph{sequentially} (within one block and across blocks).
The order-execute architecture is illustrated by~Fig.~\ref{fig:order-exec}.

Existing permissioned blockchains such as Tendermint, Chain, or Quorum typically use BFT consensus~\cite{CachinV17}, provided by PBFT~\cite{Castro:2002:PBF} or other protocols for atomic broadcast. Nevertheless, they all follow the same order-execute approach and
implement classical \emph{active} SMR~\cite{Schneider90,CharronBostPS10}.

\subsection{Limitations of Order-Execute}

The order-execute architecture is conceptually simple and therefore also widely used.  However, it has several drawbacks when used in a general-purpose permissioned blockchain.  We discuss the three most significant ones next.

\paragraph{Sequential execution.} Executing the transactions sequentially on all peers limits the effective throughput that can be achieved by the blockchain. In particular, since the throughput is inversely proportional to the execution latency, this may become a performance bottleneck for all but the simplest smart contracts. Moreover, recall that in contrast to traditional SMR, 
the blockchain forms a universal computing engine and its payload applications might be deployed by an adversary.  A denial-of-service (DoS) attack, which severely reduces the performance of such a blockchain, could simply introduce smart contracts that take a very long time to execute.  For example, a smart contract that executes an infinite loop has a fatal effect, but cannot be detected automatically because the halting problem is unsolvable.
	
To cope with this issue, public programmable blockchains with a cryptocurrency account for the execution cost.  Ethereum \cite{wood2014ethereum}, for example, introduces the concept of \emph{gas}  consumed by a transaction execution, which is converted at a \emph{gas price} to a cost in the cryptocurrency and billed to the submitter of the transaction.  Ethereum goes a long way to support this concept, assigns a cost to every low-level computation step, introducing its own VM for controlling execution.  Although this appears to be a viable solution for public blockchains, it is not adequate in the permissioned model for a general-purpose system without a native cryptocurrency.  

The distributed-systems literature proposes many ways to improve performance compared to sequential execution, for instance through parallel execution of unrelated operations~\cite{PedoneS02}.  Unfortunately, such techniques are still to be applied successfully in the blockchain context of smart contracts.  For instance, one challenge is the requirement for deterministically inferring all dependencies across smart contracts, which is particularly challenging when combined with possible confidentiality constraints.  Furthermore, these techniques are of no help against DoS attacks by contract code from untrusted developers. 

\paragraph{Non-deterministic code.} Another important problem for an order-execute architecture are \nondet transactions.  Operations executed after consensus in active SMR must be deterministic, or the distributed ledger ``forks'' and violates the basic premise of a blockchain, that all peers hold the same state.  This is usually addressed by programming blockchains in domain-specific languages (e.g., Ethereum Solidity) that are expressive enough for their applications but limited to deterministic execution.  However, such languages are difficult to design for the implementer and require additional learning by the programmer.  Writing smart contracts in a general-purpose language (e.g., Go, Java, C/C++) instead appears more attractive and accelerates the adoption of blockchain solutions.

Unfortunately, generic languages pose many problems for ensuring deterministic execution.  Even if the application developer does not introduce obviously \nondet operations, hidden implementation details can have the same devastating effect (e.g., a map iterator is not deterministic in Go).  To make matters worse, on a blockchain the burden to create deterministic applications lies on the potentially untrusted programmer. Only one \nondet contract created with malicious intent is enough to bring the whole blockchain to a halt.  A modular solution to filter diverging operations on a blockchain has also been investigated ~\cite{CachinSV16}, but it appears costly in practice.

\paragraph{Confidentiality of execution.} According to the blueprint of public blockchains, many permissioned systems run all smart contracts on all peers.  However, many intended use cases for permissioned blockchains require \emph{confidentiality}, i.e., that access to smart-contract logic, transaction data, or ledger state can be restricted.  Although cryptographic techniques, ranging from data encryption to advanced zero-knowledge proofs~\cite{Ben-SassonCG0MTV14} and verifiable computation~\cite{KosbaMSZ16}, can help to achieve confidentiality, this often comes with a considerable overhead and is not viable in practice.

Fortunately, it suffices to \emph{propagate the same state} to all peers instead of running the same code everywhere.  Thus, the execution of a smart contract can be restricted to a subset of the peers trusted for this task, that vouch for the results of the execution.  This design departs from active replication towards a variant of \emph{passive} replication~\cite{Budhiraja:1993:PA}, adapted to the trust model of blockchain.

\subsection{Further Limitations of Existing Architectures}
	
\paragraph{Fixed trust model.} Most permissioned blockchains rely on asynchronous BFT replication protocols to establish consensus~\cite{Vukolic15}.  Such protocols typically rely on a security assumption that among $n > 3f$ peers, up to $f$ are tolerated to misbehave and exhibit so-called \emph{Byzantine faults}~\cite{BrachaT85}.  The same peers often execute the applications as well, under the same security assumption (even though one could actually restrict BFT execution to fewer peers~\cite{YinMVAD03}).  However, such a quantitative trust assumption, 
irrespective of peers' roles in the system, may not match the trust required for smart-contract execution.  In a flexible system, trust at the application level should not be fixed to trust at the protocol level.  A general-purpose blockchain should decouple these two assumptions and permit flexible trust models for applications.

\paragraph{Hard-coded consensus.} Fabric is the first blockchain system that introduced pluggable consensus. Before Fabric, virtually all blockchain systems, permissioned or not, came with a hard-coded consensus protocol.
However, decades of research on consensus protocols have shown there is no such ``one-size-fits-all'' solution.  For instance, BFT protocols differ widely in their performance when deployed in potentially adversarial environments~\cite{SinghDMDR08}.  A protocol with a ``chain'' communication pattern exhibits provably optimal throughput on a LAN cluster with symmetric and homogeneous links~\cite{GuerraouiLPQ10}, 
but degrades badly on a wide-area, heterogeneous network.  Furthermore, external conditions such as load, network parameters, and actual faults or attacks may vary over time in a given deployment.  For these reasons, BFT consensus should be inherently reconfigurable and ideally adapt dynamically to a changing environment~\cite{Aublin:2015:NBP:2723895.2658994}.  Another important aspect is to match the protocol's trust assumption to a given blockchain deployment scenario. Indeed, one may want to replace BFT consensus with a protocol based on an alternative trust model such as XFT~\cite{LiuVCQV16}, or a CFT protocol, such as Paxos/Raft~\cite{Ongaro:2014:SUC:2643634.2643666} and ZooKeeper~\cite{Junqueira:2011:ZHB:2056308.2056409}, or even a permissionless protocol.

\subsection{Experience with Order-Execute Blockchain}

Prior to realizing the execute-order-validate architecture of Fabric, we gained experience with building a permissioned blockchain platform in the order-execute model, with PBFT~\cite{Castro:2002:PBF} for consensus. Namely, previous versions of \HLF (up to v0.6, released in September 2016) have been architected following the `traditional` order-execute architecture. 

From feedback obtained in many proof-of-concept applications, the limitations of this approach became immediately clear.  For instance, users often observed diverging states at the peers and reported a bug in the consensus protocol; in \emph{all cases}, closer inspection revealed that the culprit was \nondet transaction code.  Other complaints addressed limited performance, e.g., ``only five transactions per second,'' until users confessed that their average transaction took 200ms to execute.  We have learned that the key properties of a blockchain system, namely consistency, security, and performance, must \emph{not} depend on the knowledge and goodwill of its users, in particular since the blockchain should run in an untrusted environment.

%% file: architecture.tex
\section{Architecture}
\label{sec:architecture}

In this section, we introduce the three-phase \emph{execute-order-validate} architecture and then explain the transaction flow.

\subsection{\HLF Overview}
\label{sec:nutshell}

\HLF is a distributed operating system for permissioned blockchains that executes distributed applications written in general-purpose programming languages (e.g., Go, Java, Node.js).  It securely tracks its execution history in an append-only replicated ledger data structure and has no cryptocurrency built in.

\begin{figure}
  \centering
  \includegraphics[width=\columnwidth]{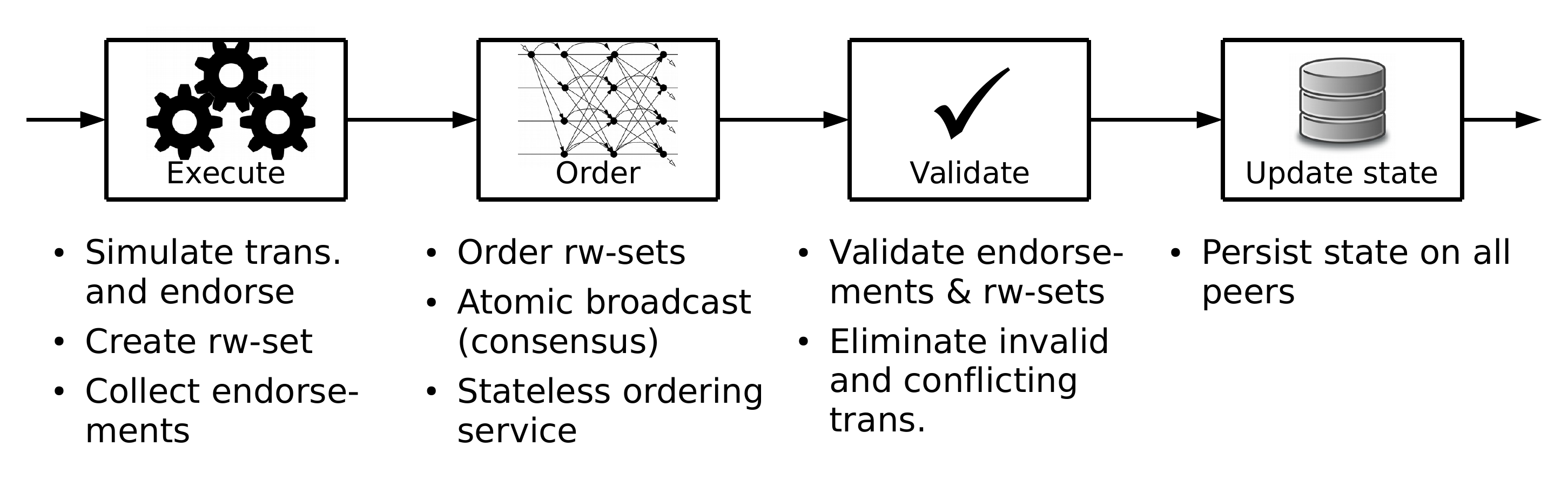}
  \caption{Execute-order-validate architecture of \HLF (\emph{rw-set} means
    a readset and writeset as explained in Sec.~\ref{sec:execution}).}
  \label{fig:exec-order-val}
\end{figure}

\HLF introduces the \emph{execute-order-validate} blockchain architecture (illustrated in Fig.~\ref{fig:exec-order-val}) and does not follow the standard order-execute design, for reasons explained in Section~\ref{sec:background}.  In a nutshell, a distributed application for \HLF consists of two parts:
\begin{itemize}
\item A smart contract, called \emph{chaincode}, which is program code that implements the application logic and runs during the \emph{execution phase}.  The chaincode is the central part of a distributed application in Fabric and may be written by an untrusted developer.  Special chaincodes exist for managing the blockchain system and maintaining parameters, collectively called \emph{system chaincodes} (Sec.~\ref{sec:chaincode}).
\item An \emph{endorsement policy} that is evaluated in the \emph{validation phase}. Endorsement policies cannot be chosen or modified by untrusted application developers.  An endorsement policy acts as a static library for transaction validation in \HLF, which can merely be parameterized by the chaincode.  Only designated \emph{administrators} may have a permission to modify endorsement policies through system management functions. A typical endorsement policy lets the chaincode specify the endorsers for a transaction in the form of a set of peers that are necessary for endorsement; it uses a monotone logical expression on sets, such as ``three out of five'' or ``$(A \land B) \lor C$.'' Custom endorsement policies may implement arbitrary logic (e.g., our Bitcoin-inspired cryptocurrency in Sec.~\ref{sec:Fabcoin}).
\end{itemize}

A client sends transactions to the peers specified by the endorsement policy.  Each transaction is then executed by specific peers and its output is recorded; this step is also called \emph{endorsement}. After execution, transactions enter the \emph{ordering phase}, which uses a pluggable consensus protocol to produce a totally ordered sequence of endorsed transactions grouped in blocks.  These are broadcast to all peers, with the (optional) help of gossip.  Unlike standard active replication~\cite{Schneider90}, which totally orders transaction \emph{inputs}, Fabric orders transaction \emph{outputs} combined with state dependencies, as computed during the execution phase.  Each peer then validates the state changes from endorsed transactions with respect to the endorsement policy and the consistency of the execution in the \emph{validation phase}.  All peers validate the transactions in the same order and validation is deterministic.  In this sense, Fabric introduces a novel \emph{hybrid replication} paradigm in the Byzantine model, which combines passive replication (the pre-consensus computation of state updates) and active replication (the post-consensus validation of execution results and state changes).

\begin{figure}
	\centering
	\includegraphics[width=\columnwidth]{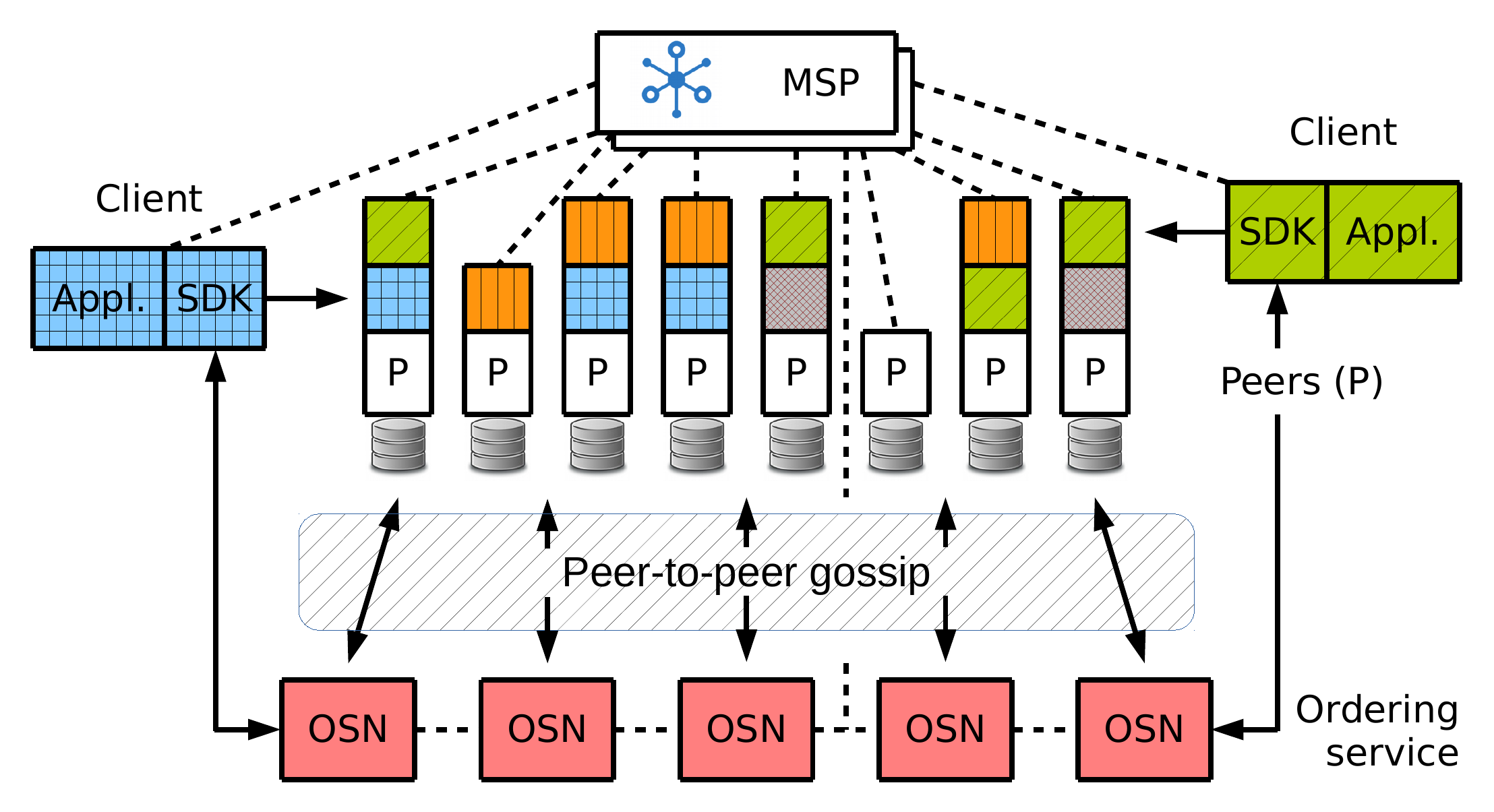}
	\caption{A \HLF network with federated MSPs and running multiple (differently shaded and colored) chaincodes, selectively installed on peers according to policy.}
	\label{fig:fabric-network}
\end{figure}

A \HLF blockchain consists of a set of \emph{nodes} that form a \emph{network} (see  Fig.~\ref{fig:fabric-network}).  As \HLF is \emph{permissioned}, all nodes that participate in the network have an identity, as provided by a modular \emph{membership service provider (MSP)} (Sec.~\ref{sec:MSP}).
Nodes in a \HLF network take up one of three roles:
\begin{itemize}
\item\emph{Clients} submit \emph{transaction proposals} for \emph{execution}, help orchestrate the execution phase, and, finally, broadcast \emph{transactions} for ordering.
\item\emph{Peers} execute transaction proposals and \emph{validate} transactions. All peers maintain the blockchain \emph{ledger}, an append-only data structure recording all transactions in the form of a hash chain, as well as the \emph{state}, a succinct representation of the latest ledger state.  Not all peers execute all transaction proposals, only a subset of them called \emph{endorsing peers} (or, simply, \emph{endorsers}) does, as specified by the policy of the chaincode to which the transaction pertains.
\item\emph{Ordering Service Nodes (OSN)} (or, simply, \emph{orderers}) are the nodes that collectively form the \emph{ordering service}. In short, the ordering service establishes the \emph{total order} of all transactions in \HLF, where each transaction contains state updates and dependencies computed during the execution phase, along with cryptographic signatures of the endorsing peers.  Orderers are entirely unaware of the application state, and do not participate in the execution nor in the validation of transactions.  This design choice renders consensus in \HLF as modular as possible and simplifies replacement of consensus protocols in \HLF.
\end{itemize}

A \HLF network actually supports multiple blockchains connected to the same ordering service.  Each such blockchain is called a \emph{channel} and may have different peers as its members.  Channels can be used to partition the state of the blockchain network, but consensus across channels is not coordinated and the total order of transactions in each channel is separate from the others.  
Certain deployments that consider all orderers as trusted may also implement by-channel access control for peers.
In the following we mention channels only briefly and concentrate on one single channel.

\begin{figure}
  \centering
  \includegraphics[width=\columnwidth]{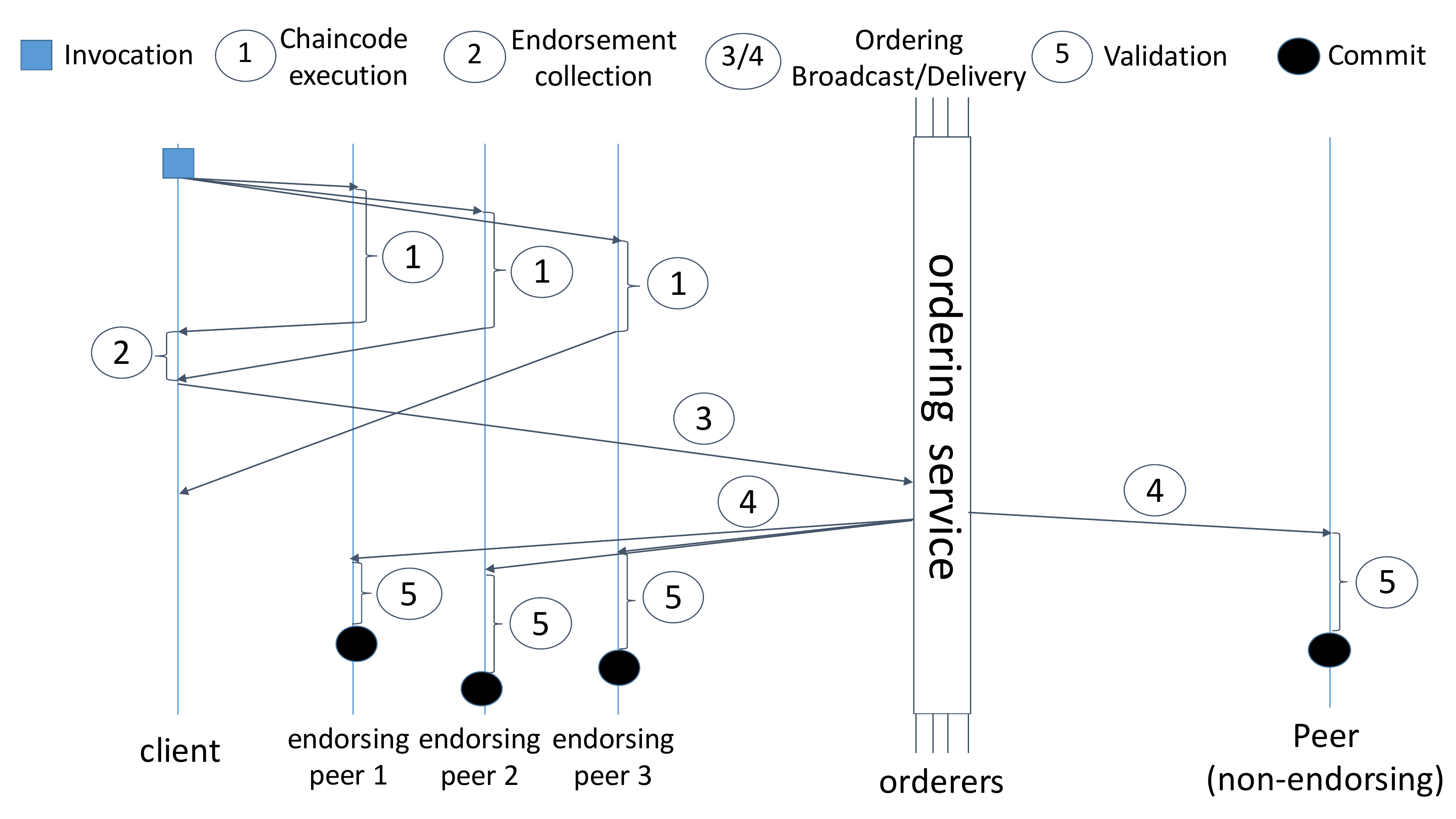}

  \caption{\HLF high level transaction flow.}
  \label{fig:flow}
\end{figure}

In the next three sections we explain the transaction flow in \HLF (depicted in Fig.~\ref{fig:flow}) and illustrate the steps of the execution, ordering, and validation phases. Then, we summarize the trust and fault model of \HLF (Sec.~\ref{sec:model}). 

\subsection{Execution Phase}
\label{sec:execution}

In the execution phase, clients sign and send the \emph{transaction proposal} (or, simply, \emph{proposal}) to one or more endorsers for execution.  Recall that every chaincode implicitly specifies a set of endorsers via the endorsement policy. A proposal contains the identity of the submitting client (according to the MSP), the transaction payload in the form of an operation to execute, parameters, and the identifier of the chaincode, a nonce to be used only once by each client (such as a counter or a random value), and a transaction identifier derived from the client identifier and the nonce.  

The endorsers \emph{simulate} the proposal, by executing the operation on the specified chaincode, which has been installed on the blockchain.
The chaincode runs in a Docker container, isolated from the main endorser process.

A proposal is simulated against the endorser's local blockchain state, without synchronization with other peers. Moreover, endorsers do not persist the results of the simulation to the ledger state. The state of the blockchain is maintained by the \emph{peer transaction manager (PTM)} in the form of a versioned key-value store, in which successive updates to a key have monotonically increasing version numbers (Sec.~\ref{sec:ledger}). The state created by a chaincode is scoped exclusively to that chaincode and cannot be accessed directly by another chaincode. Note that the chaincode is not supposed to maintain the local state in the program code, only what it maintains in the blockchain state that is accessed with \op{GetState}, \op{PutState}, and \op{DelState} operations.
Given the appropriate permission, a chaincode may invoke another chaincode to access its state within the same channel.

As a result of the simulation, each endorser produces a value \var{writeset}, consisting of the state updates produced by simulation (i.e., the modified keys along with their new values), as well as a \var{readset}, representing the version dependencies of the proposal simulation (i.e., all keys read during simulation along with their version numbers). After the simulation, the endorser cryptographically signs a message called \emph{endorsement}, which contains $\var{readset}$ and $\var{writeset}$ (together with metadata such as transaction ID, endorser ID, and endorser signature) and sends it back to the client in a \emph{proposal response}. The client collects endorsements until they satisfy the endorsement policy of the chaincode, which the transaction invokes (see Sec.~\ref{sec:validation}). In particular, this requires all endorsers as determined by the policy to produce the same execution result (i.e., identical \var{readset} and \var{writeset}). Then, the client proceeds to create the transaction and passes it to the ordering service.

\paragraph{Discussion on design choices.} As the endorsers simulate the proposal without synchronizing with other endorsers, two endorsers may execute it on different states of the ledger and produce different outputs. For the standard endorsement policy which requires multiple endorsers to produce the same result, this implies that under high contention of operations accessing the same keys, a client may not be able to  satisfy the endorsement policy. This is a new consideration compared to primary-backup replication in replicated databases with synchronization through middleware~\cite{KemmeA00}: a consequence of the assumption that no single peer is trusted for correct execution in a blockchain.

We consciously adopted this design, as it considerably simplifies the architecture and is adequate for typical blockchain applications.  As demonstrated by the approach of Bitcoin, distributed applications can be formulated such that contention by operations accessing the same state can be reduced, or eliminated completely in the normal case (e.g., in Bitcoin, two operations that modify the same ``object'' are not allowed and represent a double-spending attack~\cite{Nakamoto:Bitcoin}). In the future, we plan to gradually enhance the liveness semantics of \HLF under contention,
in particular to support CRDTs \cite{Shapiro:2011:CRD:2050613.2050642} for complementing the current version dependency checks, as well as a per-chaincode lead-endorser that would act as a transaction sequencer.

Executing a transaction before the ordering phase is critical to tolerating \nondet chaincodes (see also Sec.~\ref{sec:background}).  A chaincode in \HLF with \nondet transactions can only endanger the liveness of its own operations, because a client might not gather a sufficient number of endorsements, for instance.  This is a fundamental advantage over order-execute architecture, where \nondet operations lead to inconsistencies in the state of the peers.

Finally, tolerating \nondet execution also addresses DoS attacks from untrusted chaincode as an endorser can simply abort an execution according to a local policy if it suspects a DoS attack.  This will not endanger the consistency of the system, and again, such unilateral abortion of execution is not possible in order-execute architectures.

\subsection{Ordering Phase}
\label{sec:ordering}

When a client has collected enough endorsements on a proposal, it assembles a \emph{transaction} and submits this to the ordering service.  The transaction contains the transaction payload (i.e., the chaincode operation including parameters), transaction metadata, and a set of endorsements.  The ordering phase establishes a total order on all submitted transactions per channel.  In other words, ordering atomically broadcasts~\cite{CachinGR11} endorsements and thereby establishes consensus on transactions, despite faulty orderers.  Moreover, the ordering service batches multiple transactions into \emph{blocks} and outputs a hash-chained sequence of blocks containing transactions.  Grouping or batching transactions into blocks improves the throughput of the broadcast protocol, which is a well-known technique used in  fault-tolerant broadcasts.

At a high level, the interface of the ordering service only supports the following two operations invoked by a peer and implicitly parameterized by a channel identifier:
\begin{itemize}
\item $\textit{broadcast}(tx)$: A client calls this operation to \emph{broadcast} an arbitrary transaction~$tx$, which usually contains the transaction payload and a signature of the client, for dissemination.
\item $B \gets \textit{deliver}(s)$: A client calls this to retrieve block $B$ with non-negative sequence number~$s$. The block contains a list of transactions $[tx_1, \dots, tx_k]$ and a hash-chain value~$h$ representing the block with sequence number~$s-1$, i.e., $B = ([tx_1, \dots, tx_k], h)$.
  As the client may call this multiple times and always returns the same
  block once it is available,
  we say the peer \emph{delivers} block~$B$ with sequence number~$s$ when
  it receives $B$ \emph{for the first} time upon invoking $\textit{deliver}(s)$.
\end{itemize}

The ordering service ensures that the \emph{delivered} blocks on one channel are totally ordered.  More specifically, ordering ensures the following safety properties for each channel:
\begin{description}
\item[Agreement:] For any two blocks~$B$ delivered with sequence number~$s$ and $B'$ delivered with~$s'$ at correct peers such that $s = s'$, it holds $B=B'$.
\item[Hash chain integrity:] If some correct peer delivers a block~$B$ with number~$s$ and another correct peer delivers block
$B' =$ $([tx_1, \dots, tx_k], h')$ with number $s+1$,
then it holds $h' = H (B)$, where $H(\cdot)$ denotes the cryptographic hash function.
\item[No skipping:] If a correct peer~$p$ delivers a block with
  number~$s > 0$
  then for each $i = 0, \dots, s - 1$, peer~$p$ has already delivered a
  block with number~$i$.
\item[No creation:] When a correct peer delivers block~$B$ with
  number~$s$, then for every $tx \in B$ some client has already 
  broadcast~$tx$.
\end{description}

For liveness, the ordering service supports at least the following ``eventual'' property:
\begin{description}
\item[Validity:] If a correct client invokes $ \textit{broadcast}(tx)$,
  then every correct peer eventually delivers a block~$B$ that
  includes~$tx$, with some sequence number.
\end{description}
However, every individual  ordering implementation is allowed to come with its own liveness and fairness guarantees with respect to client requests.

Since there may be a large number of peers in the blockchain network, but only relatively few nodes are expected to implement the ordering service, \HLF can be configured to use a built-in \emph{gossip service} for disseminating delivered blocks from the ordering service to all peers (Sec.~\ref{sec:gossip}).  The implementation of gossip is scalable and agnostic to the particular implementation of the ordering service, hence it works with both CFT and BFT ordering services, ensuring the modularity of~\HLF.

The ordering service may also perform access control checks to see if a client is allowed to broadcast messages or receive blocks on a given channel.  This and other features of the ordering service are further explained in Section~\ref{sec:component.ordering}.

\paragraph{Discussion on design choices.} It is very important that the ordering service does not maintain any state of the blockchain, and neither validates nor executes transactions.  This architecture is a crucial, defining feature of \HLF, and makes \HLF the first blockchain system to totally separate consensus from execution and validation. This makes consensus as modular as possible, and enables an ecosystem of consensus protocols implementing the ordering service. The \emph{hash chain integrity} property and the chaining of blocks exist only to make the integrity verification of the block sequence by the peers more efficient. Finally, note that we do not require the ordering service to prevent transaction duplication. This simplifies its implementation and is not a concern since duplicated transactions are filtered in the read-write check by the peers during validation.

\subsection{Validation Phase}
\label{sec:validation}

Blocks are delivered to peers either directly by the ordering service or through gossip.  A new block then enters the validation phase which consists of three sequential steps:
\begin{enumerate}
\item The \emph{endorsement policy evaluation} occurs in parallel for all transactions within the block. The evaluation is the task of the so-called \emph{validation system chaincode (VSCC)}, a static library that is part of the blockchain's configuration and is responsible for validating the endorsement with respect to the endorsement policy configured for the chaincode (see Sec.~\ref{sec:chaincode}).  If the endorsement is not satisfied, the transaction is marked as invalid and its effects are disregarded.
\item A \emph{read-write conflict check} is done for all transactions in the block sequentially.  For each transaction it compares the versions of the keys in the \var{readset} field to those in the current state of the ledger, as stored locally by the peer, and ensures they are still the same.  If the versions do not match, the transaction is marked as invalid and its effects are disregarded.
\item The \emph{ledger update phase} runs last, in which the block is appended to the locally stored ledger and the blockchain state is updated.  In particular, when adding the block to the ledger, the results of the validity checks in the first two steps are persisted as well, in the form of a bit mask denoting the transactions that are valid within the block.  This facilitates the reconstruction of the state at a later time.  Furthermore, all state updates are applied by writing all key-value pairs in \var{writeset} to the local state.
\end{enumerate}

The default VSCC in \HLF allows  monotone logical expressions over the set of endorsers configured for a chaincode to be expressed.  The VSCC evaluation verifies that the set of peers, as expressed through valid signatures on endorsements of the transaction, satisfy the expression.  Different VSCC policies can be configured statically, however.

\paragraph{Discussion on design choices.} The ledger of \HLF contains all transactions, including those that are deemed invalid.  This follows from the overall design, because ordering service, which is agnostic to chaincode state, produces the chain of the blocks and because the validation is done by the peers post-consensus. This feature is needed in certain use cases that require tracking of invalid transactions during subsequent audits, and stands in contrast to other blockchains (e.g., Bitcoin and Ethereum), where the ledger contains only valid transactions. In addition, due to the permissioned nature of Fabric, detecting clients that try to mount a DoS attack by flooding the network with invalid transactions is easy. One approach would be to black-list such clients according to a policy that could be put in place. Furthermore, a specific deployment could implement transaction fees (using our currency implementation from Sec.~\ref{sec:Fabcoin} or another approach) to charge for transaction invocation, which would render a DoS attack prohibitively expensive.

\subsection{Trust and Fault Model}
\label{sec:model}

\HLF can accommodate flexible trust and fault assumptions. In general, any client is considered potentially malicious or \emph{Byzantine}. Peers are grouped into \emph{organizations} and every organization forms one trust domain, such that a peer trusts all peers within its organization but no peer of another organization. The ordering service considers all peers (and clients) as potentially Byzantine. 

The integrity of a \HLF network relies on the consistency of the ordering service. The trust model of the ordering service depends directly on its implementation (see Sec.~\ref{sec:ordering}). As of release v1.0.6, \HLF supports a centralized, single-node implementation, used in development and testing, and a CFT ordering service running on a cluster. A third implementation, a proof of concept based on \emph{BFT-SMaRt}~\cite{BessaniSA14}, tolerates up to one third of Byzantine OSNs~\cite{SousaBV18}.  

Note that \HLF decouples the trust model for applications from the trust model of consensus. Namely, a distributed application can define its own trust assumptions, which are conveyed through the endorsement policy, and are independent from those of consensus implemented by the ordering service (see also Sec.~\ref{sec:validation}).

%% file: components.tex
\section{\HLF Components}
\label{sec:components}

\begin{figure}[h]
  \vspace*{-3mm}
  \centering
  \includegraphics[width=0.8\columnwidth]{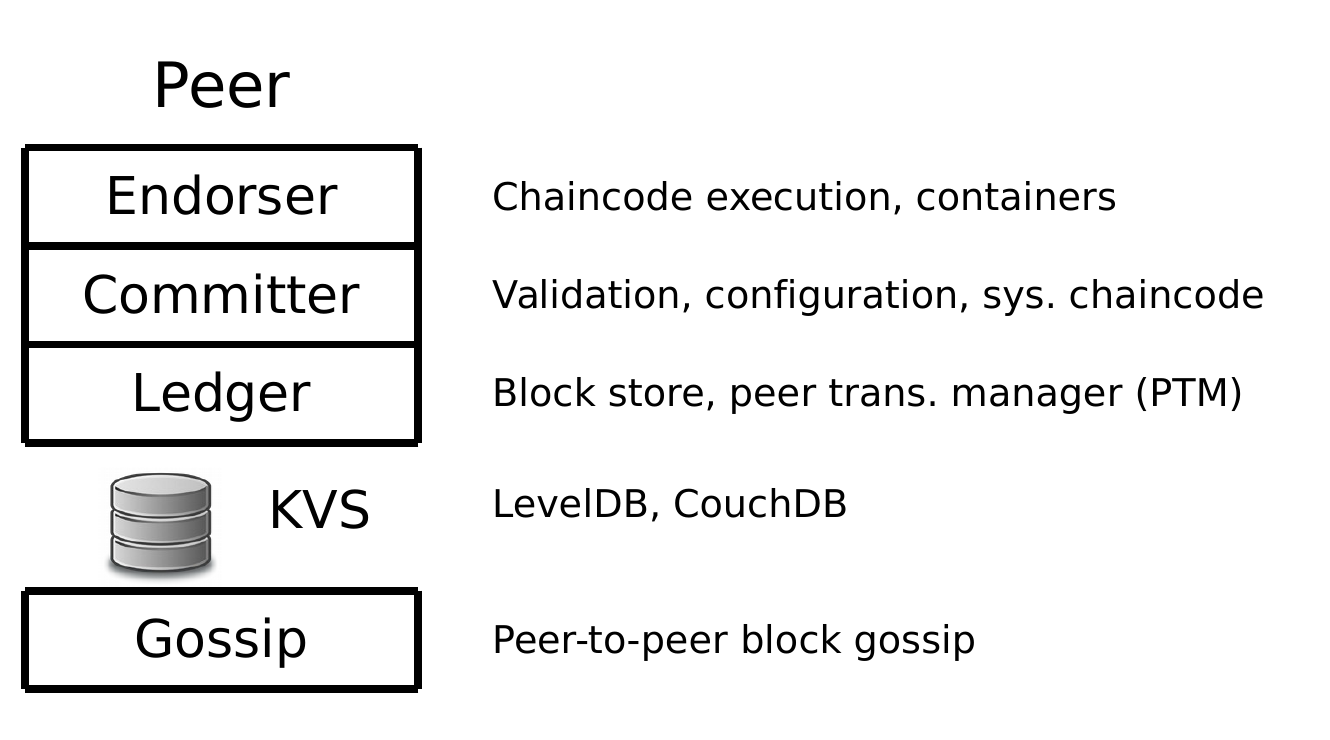}
  \caption{Components of a \HLF peer.}
  \label{fig:peer}
\end{figure}

\HLF is written in Go and uses the gRPC framework \cite{gRPCURL}
for communication between clients, peers, and orderers.  In the following we describe some important components in more detail.
Figure~\ref{fig:peer} shows the components of a peer.

\input{msp-desc}

\input{ordering}

\input{gossip}

\input{ledger}
\input{chaincodeexecution}
\input{chaincode}

%% file: msp-desc.tex
\subsection{Membership Service}
\label{sec:MSP}

The \emph{membership service provider (MSP)} maintains the identities of all nodes in the system (clients, peers, and OSNs) and is responsible for issuing node credentials that are used for authentication and authorization.  Since \HLF is \emph{permissioned}, all interactions among nodes occur through messages that are authenticated, typically with digital signatures.  The membership service comprises a component at each node, where it may authenticate transactions, verify the integrity of transactions, sign and validate endorsements, and authenticate other blockchain operations.  Tools for key management and registration of nodes are also part of the MSP. 

The MSP is an abstraction for which different instantiations are possible. The default MSP implementation in \HLF handles standard PKI methods for authentication based on digital signatures and can accommodate commercial certification authorities (CAs).  A stand-alone CA is provided as well with \HLF, called \emph{Fabric-CA}.  Furthermore, alternative MSP implementations are envisaged, such as one relying on anonymous credentials for authorizing a client to invoke a transaction without linking this to an identity~\cite{CamenischV02}.

Fabric allows two modes for setting up a blockchain network.  In \emph{offline mode}, credentials are generated by a CA and distributed out-of-band to all nodes.  Peers and orderers can only be registered in offline mode.  For enrolling clients, Fabric-CA provides an \emph{online mode} that issues cryptographic credentials to them.  The MSP configuration must ensure that all nodes, especially all peers, recognize the same identities and authentications as valid.

The MSP permits identity federation, for example, when multiple organizations operate a blockchain network.  Each organization issues identities to its own members and every peer recognizes members of all organizations.  This can be achieved with multiple MSP instantiations, for example, by creating a mapping between each organization and an MSP.

%% file: ordering.tex
\subsection{Ordering Service}
\label{sec:component.ordering}

The ordering service manages multiple channels.  On every channel, it provides the following services:
\begin{enumerate}
\item \emph{Atomic broadcast} for establishing order on transactions,
  implementing the \op{broadcast} and \op{deliver} calls (Sec.~\ref{sec:ordering}).
\item \emph{Reconfiguration} of a channel, when its members modify the
  channel by broadcasting a \emph{configuration update transaction} (Sec.~\ref{sec:chaincode}).
\item Optionally, \emph{access control}, in those configurations where the
  ordering service acts as a trusted entity, restricting broadcasting of
  transactions and receiving of blocks to specified clients and peers.
\end{enumerate}

The ordering service is bootstrapped with a \emph{genesis block} on the \emph{system channel}. This block carries a \emph{configuration transaction} that defines the  ordering service poroperties.

The current production implementation consists of \emph{or\-der\-ing-ser\-vice nodes (OSNs)} that implement the operations described here and communicate through the system channel. The actual atomic broadcast function is provided by an instance of Apache Kafka \cite{KafkaURL}
which offers scalable pub\-lish-sub\-scribe messaging and strong consistency despite node crashes, based on ZooKeeper. Kafka may run on physical nodes separate from the OSNs. The OSNs act as proxies between the peers and Kafka.

An OSN directly injects a newly received transaction to the atomic broadcast (e.g., to the Kafka broker).  OSNs \emph{batch} transactions received from the atomic broadcast and \emph{form blocks}. A block is \emph{cut} as soon as one of three conditions is met: (1) the block contains the specified maximal number of transactions; (2) the block has reached a maximal size (in bytes); or (3) an amount of time has elapsed since the first transaction of a new block was received, as explained below.

This batching process is \emph{deterministic} and therefore produces the same blocks at all nodes.  It is easy to see that the first two conditions are trivially deterministic, given the stream of transactions received from the atomic broadcast.  To ensure deterministic block production in the third case, a node starts a timer when it reads the first transaction in a block from the atomic broadcast.
If the block is not yet cut when the timer expires, the OSN broadcasts a special \emph{time-to-cut} transaction on the channel, which indicates the sequence number of the block which it intends to cut.  On the other hand, every OSN immediately cuts a new block upon receiving the \emph{first} time-to-cut  transaction for the given block number.  Since this transaction is atomically delivered to all connected OSNs, they all include the same list of transactions in the block.  
The OSNs persist a range of the most recently delivered blocks directly to their filesystem, so they can answer to peers retrieving blocks through \emph{deliver}.

The ordering service based on Kafka is one of three implementations currently available.  A centralized orderer, called \emph{Solo}, runs on one node and is used for development.  A proof-of-concept ordering service based on \emph{BFT-SMaRt}~\cite{BessaniSA14} has also been made available~\cite{SousaBV18}; it ensures the atomic broadcast service, but not yet reconfiguration and access control.  This illustrates the modularity of consensus in Fabric.

%% file: gossip.tex
\subsection{Peer Gossip}
\label{sec:gossip}

One advantage of separating the execution, ordering, and validation phases is that they can be scaled independently.  However, since most consensus algorithms (in the CFT and BFT models) are bandwidth-bound, the throughput of the ordering service is capped by the network capacity of its nodes. Consensus cannot be scaled up by adding more nodes~\cite{croman2016scaling, Vukolic15}, rather, throughput will decrease.  However, since ordering and validation are decoupled, we are interested in efficiently broadcasting the execution results to all peers for validation, after the ordering phase.  This, along with \emph{state transfer} to newly joining peers and peers that were disconnected for a long time, is precisely the goal of the gossip component. \HLF gossip utilizes  epidemic multicast~\cite{demers1987epidemic} for this purpose.  The blocks are signed by the ordering service.  This means that a peer can, upon receiving all blocks,   independently assemble the blockchain and verify its integrity.

The communication layer for gossip is based on gRPC and utilizes TLS with mutual authentication, which enables each side to bind the TLS credentials to the identity of the remote peer.
The gossip component maintains an up-to-date \emph{membership view} of the online peers in the system.  All peers independently build a local view from periodically disseminated membership data. Furthermore, a peer can reconnect to the view after a crash or a network outage.

\HLF gossip uses two phases for information dissemination: during \emph{push}, each peer selects a random set of active neighbors from the membership view, and forwards them the message; during \emph{pull}, each peer periodically probes a set of randomly selected peers and requests missing messages. It has been shown~\cite{demers1987epidemic, karp2000randomized} that using both methods in tandem is crucial to optimally utilize the available bandwidth and to ensure that all peers receive all messages with high probability.  In order to reduce the load of sending blocks from the ordering nodes to the network, the protocol also \emph{elects a leader peer} that pulls blocks from the ordering service on their behalf and initiates the gossip distribution.  This mechanism is resilient to leader failures.

%% file: ledger.tex
\subsection{Ledger}
\label{sec:ledger}

The ledger component at each peer maintains the ledger and the state on persistent storage and enables \emph{simulation}, \emph{validation}, and \emph{ledger-update} phases. Broadly, it consists of a \emph{block store} and a \emph{peer transaction manager}. 

The \emph{ledger block store} persists transaction blocks and is implemented as a set of append-only files. Since the blocks are immutable and arrive in a definite order, an append-only structure gives maximum performance. In addition, the block store maintains a few indices for random access to a block or to a transaction in a block.

The \emph{peer transaction manager (PTM)} maintains the \emph{latest state} in a versioned key-value store. It stores one tuple of the form $(\var{key}, \var{val}, \var{ver})$ for each unique entry \var{key} stored by any chaincode, containing its most recently stored value \var{val} and its latest version \var{ver}. The version consists of the block sequence number and the sequence number of the transaction (that stores the entry) within the block. This makes the version unique and monotonically increasing. The PTM uses a local key-value store to realize its versioned variant, with implementations using LevelDB (in Go) \cite{LevelDBGoURL}
and Apache CouchDB \cite{CouchDBGoURL}.

During simulation the PTM provides a stable snapshot of the latest state to the transaction.  As mentioned in Section~\ref{sec:execution}, the PTM records in \var{readset} a tuple $(\var{key}, \var{ver})$ for each entry accessed by \op{GetState} and in \var{writeset} a tuple $(\var{key}, \var{val})$ for each entry updated with \op{PutState} by the transaction. In addition, the PTM supports range queries, for which it computes a cryptographic hash of the query results (a set of tuples $(\var{key}, \var{ver})$) and adds the query string itself and the hash to \var{readset}.

For transaction validation (Sec.~\ref{sec:validation}), the PTM validates all transactions in a block sequentially. This checks whether a transaction conflicts with any preceding transaction (within the block or earlier). For any key in \var{readset}, if the version recorded in \var{readset} differs from the version present in the latest state (assuming that all preceding valid transactions are committed), then the PTM marks the transaction as invalid. For range queries, the PTM re-executes the query and compares the hash with the one present in \var{readset}, to ensure that no phantom reads occur.  This read-write conflict semantics results in one-copy serializability~\cite{Kemme09}. 

The ledger component tolerates a crash of the peer during the ledger update as follows.  After receiving a new block, the PTM has already performed validation and marked transactions as valid or invalid within the block, using a bit mask as mentioned in Section~\ref{sec:validation}.  The ledger now writes the block to the ledger block store, flushes it to disk, and subsequently updates the block store indices.  Then the PTM applies the state changes from \var{writeset} of all valid transactions to the local versioned store.  Finally, it computes and persists a value \var{savepoint}, which denotes the largest successfully applied block number.  The value \var{savepoint} is used to recover the indices and the latest state from the persisted blocks when recovering from a crash.

%% file: chaincodeexecution.tex
\subsection{Chaincode Execution}
\label{sec:chaincode-execution}

Chaincode is executed within an environment loosely coupled to the rest of the peer, which supports plugins for adding new chaincode programming languages. Currently Go, Java, and Node are supported.

Every user-level or application chaincode runs in a separate process within a Docker container environment, which isolates the chaincodes from each other and from the peer. This also simplifies the management of the lifecycle for chaincodes (i.e., starting, stopping, or aborting chaincode). The chaincode and the peer communicate using gRPC messages. Through this loose coupling, the peer is agnostic of the actual language in which chaincode is implemented.

In contrast to application chaincode, system chaincode runs directly in the peer process.  System chaincode can implement specific functions needed by \HLF and may be used in situations where the isolation among user chaincodes is overly restrictive. 
More details on system chaincodes are given in the next section.

%% file: chaincode.tex
\subsection{Configuration and System Chaincodes}
\label{sec:chaincode}

\HLF is customized through \emph{channel configuration} and through special chaincodes, known as \emph{system chaincodes}.

Recall each channel in \HLF forms one logical blockchain.
The \emph{configuration} of a channel is maintained in metadata persisted in special \emph{configuration blocks}.
Each configuration block contains the full channel configuration and does not contain any other transactions.  Each blockchain begins with a configuration block known as the \emph{genesis block} which is used to bootstrap the channel.  The channel configuration includes: (1) definitions of the MSPs for the participating nodes, (2) the network addresses of the OSNs, 
(3) shared configuration for the consensus implementation and the ordering service, such as batch size and timeouts, (4) rules governing access to the ordering service operations (\textit{broadcast}, and \textit{deliver}), and (5) rules governing how each part the channel configuration may be modified.

The configuration of a channel may be updated using a \emph{channel configuration update transaction}.  This transaction contains a representation of the changes to be made to the configuration, as well as a set of signatures.  The ordering service nodes evaluate whether the update is valid by using the current configuration to verify that the modifications are authorized using the signatures.  The orderers then generate a new configuration block, which embeds the new configuration and the configuration update transaction.  Peers receiving this block validate whether the configuration update is authorized based on the current configuration; if valid, they update their current configuration.

The application chaincodes are deployed with a reference to an \emph{endorsement system chaincode (ESCC)} and to a \emph{validation system chaincode (VSCC)}.  These two chaincodes are selected such that the output of the ESCC (an endorsement) may be validated as part of the input to the VSCC.  The ESCC takes as input a proposal and the proposal simulation results.  If the results are satisfactory, then the ESCC produces a response, containing the results and the endorsement.  For the default ESCC, this endorsement is simply a signature by the peer's local signing identity.  The VSCC takes as input a transaction and outputs whether that transaction is valid.  For the default VSCC, the endorsements are collected and evaluated against the endorsement policy specified for the chaincode.
Further system chaincodes implement other support functions, such as chaincode lifecycle.

%% file: evaluation.tex
\section{Evaluation}
\label{sec:evaluation}

Even though \HLF is not yet performance-tuned and optimized, we report in this section on some preliminary performance numbers. \HLF is a complex distributed system; its performance depends on many parameters including the choice of a distributed application and transaction size, the ordering service and consensus implementation and their parameters, the network parameters and topology of nodes in the network, the hardware on which nodes run, the number of nodes and channels, further configuration parameters, and the network dynamics. Therefore, in-depth performance evaluation of \HLF is postponed to future work.

In the absence of a standard benchmark for blockchains, we use the most prominent blockchain application for evaluating \HLF, a simple authority-minted cryptocurrency that uses the data model of Bitcoin, which we call \emph{Fabric coin} (abbreviated hereafter as \emph{\coin}). This allows us to  put the performance of \HLF in the context of other permissioned blockchains, which are often derived from Bitcoin or Ethereum.  For example, it is also the application used in benchmarks of other permissioned blockchains~\cite{SettyBZRV17,quorum17}.

In the following, we first describe \coin (Sec.~\ref{sec:Fabcoin}), which also demonstrates how to customize the validation phase and endorsement policy.  In Section~\ref{sec:results} we present the benchmark and discuss our results.

\subsection{Fabric Coin (\coin)}
\label{sec:Fabcoin}

\paragraph{UTXO cryptocurrencies} The data model introduced by Bitcoin~\cite{Nakamoto:Bitcoin} has become known as ``unspent transaction output'' or \emph{UTXO}, and is also used by many other cryptocurrencies and distributed applications.
UTXO represents each step in the evolution of a data object as a separate atomic state on the ledger. Such a state is created by a transaction and destroyed (or ``consumed'') by another unique transaction occurring later. Every given transaction destroys a number of \emph{input states} and creates one or more \emph{output states}.
A ``coin'' in Bitcoin is initially created by a \emph{coinbase} transaction that rewards the ``miner'' of the block. This appears on the ledger as a \emph{coin state} designating the miner as the owner. Any coin can be \emph{spent} in the sense that the coin is assigned to a new owner by a transaction that atomically destroys the current coin state designating the previous owner and creates another coin state representing the new owner.

We capture the UTXO model in the key-value store of \HLF as follows. Each UTXO state corresponds to a unique KVS entry that is created once (the coin state is ``unspent'') and destroyed once (the coin state is ``spent''). Equivalently, every state may be seen as a KVS entry with logical version~$0$ after creation; when it is destroyed again, it receives version~$1$. There should not be any concurrent updates to such entries (e.g., attempting to update a coin state in different ways amounts to double-spending the coin).

Value in the UTXO model is transferred through transactions that refer to several input states that all belong to the entity issuing the transaction.  An entity owns a state because the public key of the entity is contained in the state itself.  Every transaction creates one or more output states in the KVS representing the new owners, deletes the input states in the KVS, and ensures that the sum of the values in the input states equals the sum of the output states' values. There is also a policy determining how value is created  (e.g., \emph{coinbase} transactions in Bitcoin or specific \emph{mint} operations in other systems) or destroyed.

\paragraph{\coin implementation}

Each state in \coin is a tuple of the form $(\var{key}, \var{val}) = (\var{txid}.j, (\var{amount}, \var{owner}, \var{label}))$, denoting the coin state created as the $j$-th output of a transaction with identifier \var{txid} and allocating \var{amount} units labeled with~\var{label} to the entity whose public key is~\var{owner}. Labels are strings used to identify a given type of a coin (e.g., `USD`, `EUR`, `FBC`). Transaction identifiers are short values that uniquely identify every \HLF transaction.
The \coin implementation consists of three parts: (1) a client wallet, (2) the \coin chaincode, and (3) a custom VSCC for \coin implementing its endorsement policy.

\paragraph{Client wallet} By default, each \HLF client maintains a \emph{\coin wallet} that locally stores a set of cryptographic keys allowing the client to spend coins. For creating a \textsc{spend} transaction that transfers one or more coins, the client wallet creates a \coin request $\var{request} = (\var{inputs}, \var{outputs}, \var{sigs})$ containing: (1) a list of \emph{input coin states}, $\var{inputs} = [\var{in}, \dots]$ that specify coin states ($\var{in}, \dots$) the client wishes to spend, as well as (2) a list of \emph{output coin states}, $\var{outputs} = [(\var{amount}, \var{owner}, \var{label}), \dots]$. The client wallet signs, with the private keys that correspond to the input coin states, the concatenation of the \coin request and a nonce, which is a part of every \HLF transaction, and adds the signatures in a set~\var{sigs}. A \textsc{spend} transaction is valid when the sum of the amounts in the input coin states is at least the sum of the amounts in the outputs and when the output amounts are positive. For a \textsc{mint} transaction that creates new coins, $\var{inputs}$ contains only an identifier (i.e., a reference to a public key) of a special entity called \emph{Central Bank (CB)}, whereas \var{outputs} contains an arbitrary number of coin states. To be considered valid, the signatures of a \textsc{mint} transaction in \var{sigs} must be a cryptographic signature under the public key of CB over the concatenation of the \coin request and the aforementioned nonce. \coin may be configured to use multiple CBs or specify a threshold number of signatures from a set of CBs. Finally, the client wallet includes the \coin request into a transaction and sends this to a peer of its choice.

\paragraph{\coin chaincode} A peer runs the chaincode of \coin which simulates the transaction and creates readsets and writesets. In a nutshell, in the case of a \textsc{spend} transaction, for every input coin state $\var{in} \in \var{inputs}$ the chaincode first performs $\op{GetState}(\var{in})$; this includes $\var{in}$ in the readset along with its current version (Sec.~\ref{sec:ledger}). Then the chaincode executes $\op{DelState}(\var{in})$ for every input state~\var{in}, which also adds \var{in} to the writeset and effectively marks the coin state as ``spent.'' Finally, for $j= 1, \dots, |\var{outputs}|$, the chaincode executes $\op{PutState}(\var{txid}.j, \var{out})$ with the $j$-th output $\var{out} = (\var{amount}, \var{owner}, \var{label})$ in $\var{outputs}$.
In addition, a peer may run the transaction validation code as described next in the VSCC step for \coin; this is not necessary, since the \coin VSCC actually validates transactions, but it allows the (correct) peers to filter out potentially malformed transactions. In our implementation, the chaincode runs the \coin VSCC without cryptographically verifying the signatures (these are verified only in the actual VSCC).

\paragraph{Custom VSCC} Finally, every peer validates \coin transactions using a custom VSCC. This verifies the cryptographic signature(s) in \var{sigs} under the respective public key(s) and performs semantic validation as follows. For a \textsc{mint} transaction, it checks that the output states are created under the matching transaction identifier (\var{txid}) and that all output amounts are positive. For a \textsc{spend} transaction, the VSCC additionally verifies (1) that for all input coin states, an entry in the readset has been created and that it was also added to the writeset and marked as deleted, (2) that the sum of the amounts for all input coin states equals the sum of amounts of all output coin states, and (3) that input and output coin labels match.  Here, the VSCC obtains the input coin amounts by retrieving their current values from the ledger.

Note that the \coin VSCC does not check transactions for double spending, as this occurs through \HLF's standard validation that runs after the custom VSCC. In particular, if two transactions attempt to assign the same unspent coin state to a new owner, both would pass the VSCC logic but would be caught subsequently in the read-write conflict check performed by the PTM.  According to Sections~\ref{sec:validation} and \ref{sec:ledger}, the PTM verifies that the current version number stored in the ledger matches the one in the readset; hence, after the first transaction has changed the version of the coin state, the transaction ordered second will be recognized as invalid.

\subsection{Experiments}
\label{sec:results}

\paragraph{Setup} Unless explicitly mentioned differently, in our experiments: (1) nodes run on Fabric version v1.1.0-preview\footnote{Patched with commit IDs $9e770062$ and $eb437dab$ in the Fabric \emph{master} branch.} instrumented for performance evaluation through local logging, (2) nodes are hosted in a single IBM Cloud (SoftLayer) data center (DC) as dedicated VMs interconnected with 1Gbps (nominal) networking, (3) all nodes are 2.0 GHz 16-vCPU VMs running Ubuntu with 8GB of RAM and SSDs as local disks, (4) a single-channel ordering service runs a typical Kafka orderer setup with 3 ZooKeeper nodes, 4 Kafka brokers and 3 \HLF orderers, all on distinct VMs, (5) there are 5 peers in total, all belonging to different organizations (orgs) and all being \coin endorsers, and (6) signatures use the default 256-bit ECDSA scheme. In order to measure and stage latencies in the transaction flow spanning multiple nodes, the node clocks are synchronized with an NTP service throughout the experiments.  All communication among \HLF nodes is configured to use TLS.

\paragraph{Methodology} In every experiment, in the first phase we invoke transactions that contain only \coin \textsc{mint} operations to produce the coins, and then run a second phase of the experiment in which we invoke \coin \textsc{spend} operation on previously minted coins (effectively running single-input, single-output \textsc{spend} transactions).  When reporting throughput measurements, we use an increasing number of \HLF CLI clients (modified to issue concurrent requests) running on a single VM, until the end-to-end throughput is saturated, and state the throughput just \emph{below} saturation. Throughput numbers are reported as the average measured during the steady state of an experiment, disregarding the ``tail,'' where some client threads already stop submitting their share of transactions. In every experiment, the client threads collectively invoke at least 500k \textsc{mint} and \textsc{spend} transactions.

\begin{figure}[!htbp]
		\centering
		\includegraphics[width=\columnwidth]{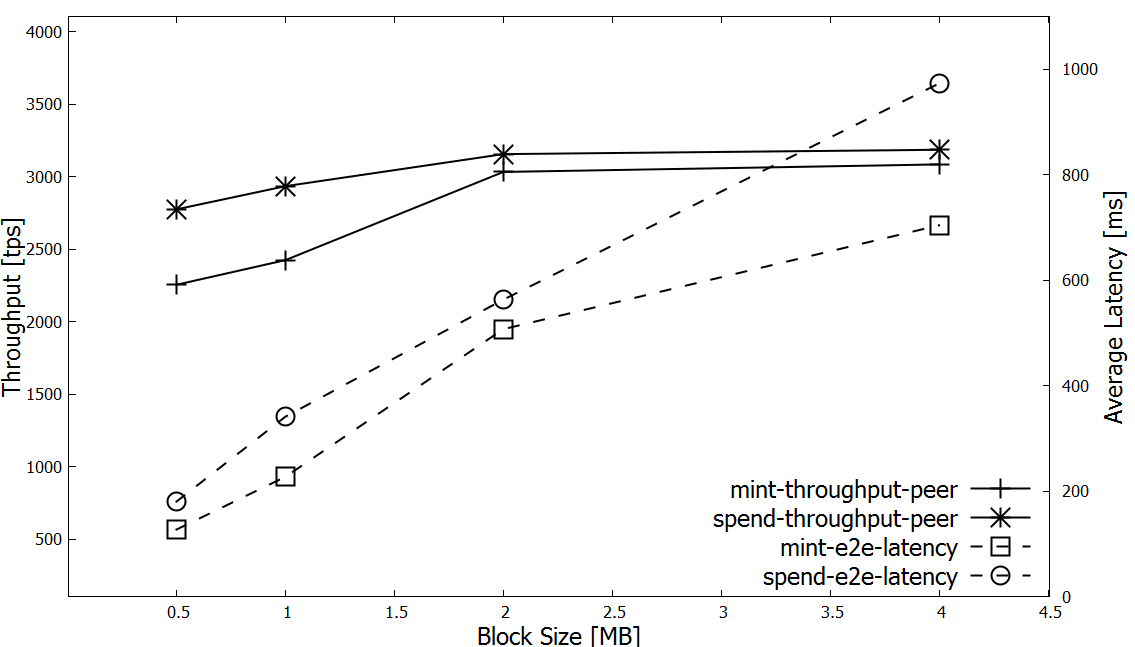}

		\caption{Impact of block size on throughput and latency.}
		\label{fig:blocksize}
	\end{figure}

\paragraph{Experiment 1: Choosing the block size} A critical \HLF configuration  parameter that impacts both throughput and latency is block size. To fix the block size for subsequent experiments, and to evaluate the impact of block size on performance, we ran experiments varying block size from 0.5MB to 4MB. Results are depicted in Fig.~\ref{fig:blocksize}, showing peak throughput measured at the peers along with the corresponding average end-to-end (e2e) latency.

We can observe that throughput does not significantly improve beyond a block size of 2MB, but latency gets worse (as expected). Therefore, we adopt 2MB as the block size for the following experiments, with the goal of maximizing the measured throughput, assuming the end-to-end latency of roughly 500ms is acceptable.

\paragraph{Size of transactions} During this experiment, we also observed the size \textsc{mint} and \textsc{spend} transactions. In particular, the 2MB-blocks contained 473 \textsc{mint} or 670 \textsc{spend} transactions, i.e., the average transaction size is 3.06kB for \textsc{spend} and 4.33kB for \textsc{mint}. In general, transactions in \HLF are large because they carry certificate information. Besides, \textsc{mint} transactions of \coin are larger than \textsc{spend} transactions because they carry CB certificates. This is an avenue for future improvement of both \HLF and \coin.

\paragraph{Experiment 2: Impact of peer CPU} Fabric peers run many CPU-intensive cryptographic operations. To estimate the impact of CPU power on throughput, we performed a set of experiments in which 4 peers run on 4, 8, 16, and 32 vCPU VMs, while also doing coarse-grained latency staging of block validation to  identify bottlenecks. 

Our experiment focused on the validation phase, as ordering with the Kafka ordering service has never been a bottleneck in our cluster experiments (within one data center). The validation phase, and in particular the VSCC validation of \coin, is computationally intensive, due to its many digital signature verifications. We calculate the validation throughput at the peer based on measuring validation phase latency locally at the peer.

	\begin{figure}[htbp]
		\centering
		\begin{subfigure}[b]{\columnwidth}
		\includegraphics[width=1\linewidth]{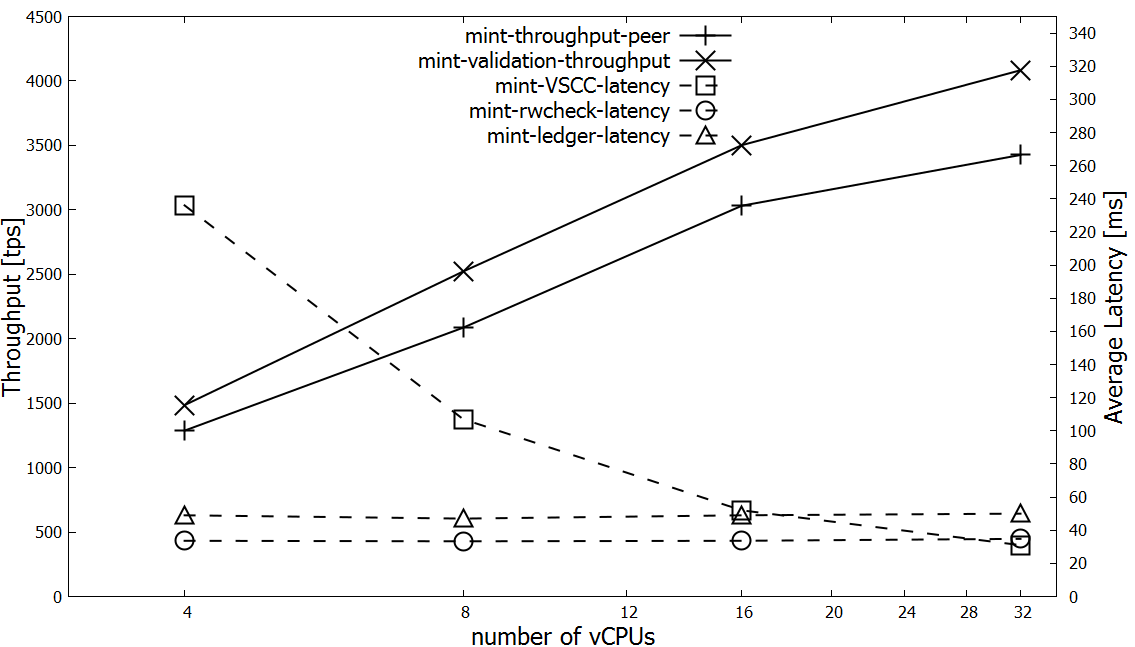}
		\caption{Blocks containing only \textsc{mint} transactions.}
		\label{fig:cpumint}
		\end{subfigure}%
		\bigbreak
		\begin{subfigure}[b]{\columnwidth}

		\includegraphics[width=1\linewidth]{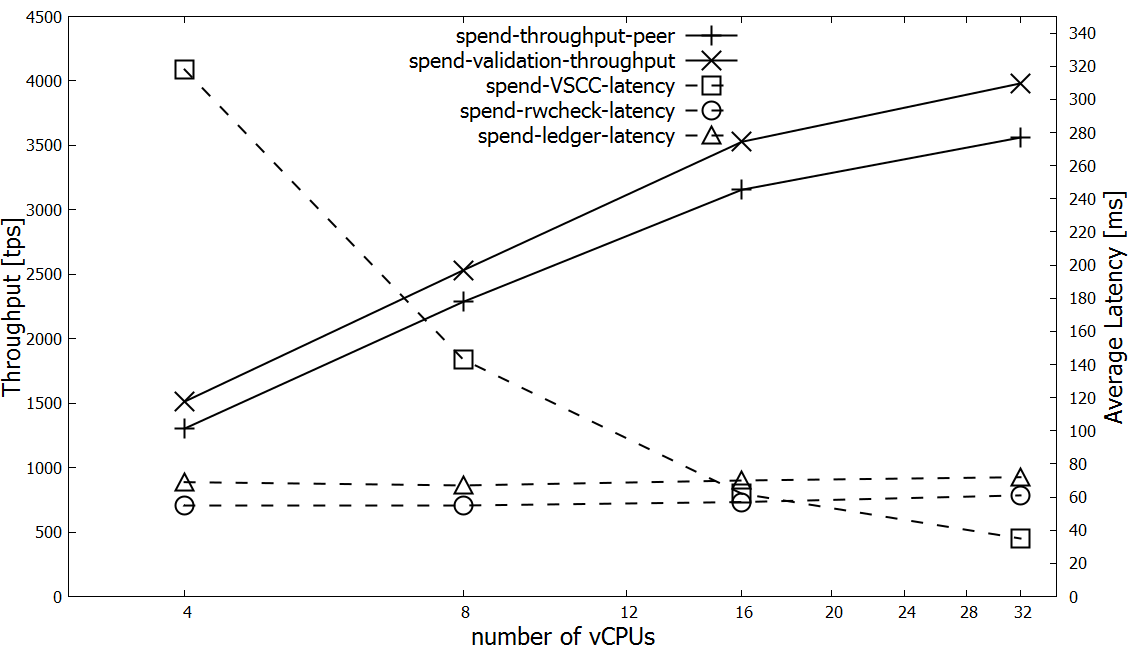}
		\caption{Blocks containing only \textsc{spend} transactions.}
		\label{fig:cpuspend}
		\end{subfigure}%

		\caption{Impact of peer CPU on end-to-end throughput, validation throughput and block validation latency.}
		\label{fig:CPU}
	\end{figure}

The results, with 2MB blocks, are shown in Fig.~\ref{fig:CPU}, for blocks containing \textsc{mint}  (Fig.~\ref{fig:cpumint}) and \textsc{spend} (Fig.~\ref{fig:cpuspend}) operations. For both operations the measured throughput and latency scale in the same way with the number of vCPUs. We can observe that the validation effort clearly limits the achievable (end-to-end) throughput. Furthermore, the validation performance by the \coin VSCC scales quasi-linearly with CPU, as the endorsement policy verification by \HLF's VSCC is embarrassingly parallel. However, the read-write-check and ledger-access stages are sequential and become dominant with a larger number of cores (vCPUs). This is in particular noticeable for \textsc{spend} transactions, since more \textsc{spend} than \textsc{mint} transactions can fit into a 2MB block, which prolongs the duration of the sequential validation stages (i.e., read-write-check and ledger-access). 

This experiment suggests that future versions of \HLF could profit from pipelining the validation stages (which are now sequential), removing sequential overhead in the peer that causes a noticeable difference between validation and end-to-end throughput, optimizing stable-storage access, and parallelizing read-write dependency checks. 

Finally, in this experiment, we measured over 3560 transactions per second (tps) average \textsc{spend} (end-to-end) throughput at the 32-vCPU peer. The \textsc{mint} throughput is, in general, slightly lower than that of \textsc{spend}, but the difference is within 10\%, with 32-vCPU peer reaching over 3420 tps average \textsc{mint} throughput.

\paragraph{Latency profiling by stages} We further performed coarse-grained profiling of latency during our previous experiment at the peak reported throughput.  Results are depicted in Table~\ref{table:latency32coresFull}. The ordering phase comprises broadcast-deliver latency and internal latency within a peer before validation starts. The table reports average latencies for \textsc{mint} and \textsc{spend}, standard deviation, and tail latencies.

\begin{table}[htbp]

	\begin{footnotesize}

		\centering

		\begin{tabular}{|l|c|c|c|c|}

			\cline{2-5}

			\multicolumn{1}{c|}{} & {avg} & {st.dev} & {99\%} & {99.9\%} \\

			\hline

			(1) endorsement &
			5.6 /\ 7.5 &
      2.4 /\ 4.2 &
      15 /\ 21	&
      19 /\ 26 \\

			\hline

			(2) ordering &

	     248 /\ 365 &
       60.0 /\ 92.0 &
       484 /\ 624	&
       523 /\  636 \\

			\hline

			(3) VSCC val. &

	     31.0 /\ 35.3	&
       10.2 /\  9.0	&
       72.7  /\ 57.0	&
       113 /\  108.4 \\

			\hline

			(4) R/W check &

			34.8 /\ 61.5	&
      3.9 /\ 9.3	&
      47.0 /\ 88.5	&
      59.0 /\ 93.3 \\

			\hline

			(5) ledger &

			50.6 /\ 72.2 	&
      6.2 /\ 8.8	&
      70.1  /\ 97.5	&
      72.5 /\ 105 \\

			\hline

			(6) validation (3+4+5)&

		116 /\ 169 &
    12.8 /\ 17.8	&
    156 /\ 216	&
    199 /\ 230 \\

			\hline

			(7) end-to-end (1+2+6) &

	     371 /\ 542	&
       63 /\ 94	&
       612 /\ 805 &
       646 /\ 813 \\

			\hline

		\end{tabular}

	\end{footnotesize}

	\caption{Latency statistics in milliseconds (ms) for \textsc{mint} and \textsc{spend}, broken down into five stages at a 32-vCPU peer with 2MB blocks.  Validation (6) comprises stages 3, 4, and~5; the end-to-end latency contains stages~1--5.}
	\label{table:latency32coresFull}

\end{table}

We observe that ordering dominates the overall latency.
We also see that average latencies lie below 550ms with sub-second tail latencies. In particular, the highest end-to-end latencies in our experiment come from the first blocks, during the load build-up. Latency under lower load can be regulated and reduced using the time-to-cut parameter of the orderer (see Sec.~\ref{sec:ordering}), which we basically do not use in our experiments, as we set it to a large value.

\paragraph{Experiment 3: SSD vs. RAM disk} To evaluate the potential gains related to stable storage, we repeated the previous experiment with RAM disks (tmpfs) mounted as stable storage at all peer VMs. The benefits are limited, as tmpfs only helps with the ledger phase of the validation at the peer. We measured sustained peak throughput at 3870 \textsc{SPEND} tps at 32-vCPU peer, roughly a 9\% improvement over SSD.

\paragraph{Experiment 4: Scalability on LAN} In this and the following experiment we increase the number of peers (with 16 vCPUs each) to evaluate the scalability of Fabric.

\begin{figure}[!htbp]
	\centering
	\includegraphics[width=\columnwidth]{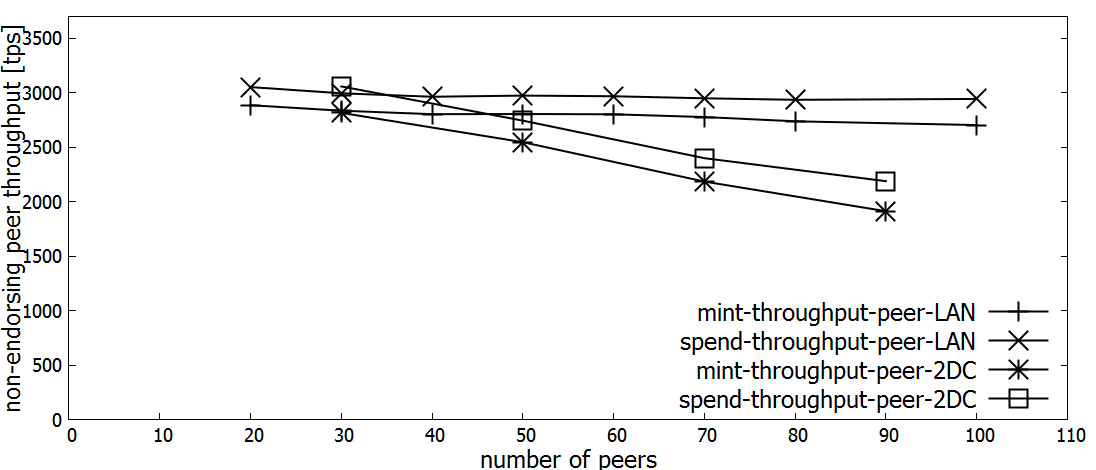}
	\caption{Impact of varying number of peers on non-endorsing peer throughput.}
	\label{fig:scaling}
\end{figure}

In this experiment we maintain one peer per organization hosted in a single IBM Cloud DC (Hong Kong, HK). All peers receive blocks directly from the ordering service without gossip. We start from 20 peers (10 of which are \coin endorsers) and increase the number of peers to 100. The achievable peak throughput in function of the number of peers is depicted in Fig.~\ref{fig:scaling} (``LAN'' suffix). 

We observe that the Kafka ordering service handles the added number of peers well and scales with the increase. As peers connect randomly to OSNs, the bandwidth of 3 OSNs should become a bottleneck with the 1Gbps nominal throughput of the network. However, this does not occur. We tracked down the reason for this and found that the provisioned bandwidth in the IBM Cloud was higher than nominal one, with \op{netperf} reporting consistently 5-6.5Gbps between pairs of nodes. 

\paragraph{Experiment 5: Scalability over two DCs and impact of gossip} In a follow-up experiment, we moved the ordering service, 10 endorsers, and the clients to the nearby Tokyo (TK) data center, leaving the non-endorsing peers in the HK DC. The goal in this (and the next) experiment is to evaluate the system when the network bandwidth becomes the bottleneck. We varied the number of non-endorsing peers in HK from 20 to 80, maintaining direct connectivity with the ordering service (i.e., one peer per org), in addition to 10 endorsing peers in TK. The single-TCP netperf throughput reported between two VMs in TK and HK is 240 Mbps on average. 

The peak throughput in function of the (total) number of peers is depicted in Fig.~\ref{fig:scaling} (``2DC'' suffix). We clearly see that the throughput is basically the same as in the previous experiment with 30 peers, but it drops when the number of peers increases. The throughput is reduced since the network connections of 3 OSNs in TK are saturated. We measured 1910 tps \textsc{mint} and 2190 tps \textsc{spend} throughput (at HK peers) with a total of 90 peers in this configuration. 

To cope with this, and to improve the scalability over a WAN, \HLF may employ gossip (Sec.~\ref{sec:gossip}). We repeated the last measurement with 80 peers in HK (totaling 90 peers) but reconfigured these peers into 8 orgs of 10 peers each. In this configuration, only one leader peer per org connects directly to the ordering service and gossips the blocks to the others in its org. This experiment (with a gossip fanout of 7) achieves 2544/2753 tps \textsc{mint/spend} average peak throughput at HK peers, which means that gossip nicely serves its intended function. The throughput is somewhat lower than in the LAN experiment, as org leader peers (directly connected to OSNs in both experiments) now need to manage gossip as well. 

\paragraph{Experiment 6: Performance over multiple data centers (WAN)} Finally, we extend the last experiment to 5 different data centers: Tokyo (TK), Hong Kong (HK), Melbourne (ML), Sydney (SD), and Oslo (OS), with 20 peers in each data center, totaling 100 peers. As in the previous experiment, the ordering service, 10 endorsers, and all clients are in TK. We run this experiment without gossip (one peer per org) and with gossip (10 orgs of 10 peers, 2 orgs per data center, fanout 7). The results are summarized in Table~\ref{table:WAN}, averaged across peers belonging to same data center. For reference, the first row of the table shows the netperf throughput between a VM in a given data center and TK.

\begin{table}[htbp]
	
	\begin{footnotesize}
		
		\centering
		
		\begin{tabular}{|c|c|c|c|c|}
			
			\cline{2-5}
			
			\multicolumn{1}{c|}{} & {HK} & {ML} & {SD} & OS \\
			
			\hline
			
			netperf to TK [Mbps] & 240 & 98 & 108 & 54 \\
			
			\hline
			\makecell{peak \textsc{mint} / \textsc{spend} \\ throughput [tps] \\(without gossip)  }&
			1914 /\ 2048  & 1914 /\ 2048 & 1914 /\ 2048 & 1389 /\ 1838 \\
			
			\hline

			\makecell{peak \textsc{mint} / \textsc{spend} \\ throughput [tps] \\(with gossip)  }&
			2553 /\ 2762  & 2558 /\ 2763 & 2271 /\ 2409 & 1484 /\ 2013 \\
			\hline
			
		\end{tabular}
		
	\end{footnotesize}
	
	\caption{Experiment with 100 peers across 5 data centers.}
	\label{table:WAN}
	
\end{table}

The results again clearly show the benefits of using gossip when the peers are scattered over a WAN. We observe interesting results with the peers in OS and SD compared to HK and ML. The lower throughput with gossip for SD is due to CPU limitations of VMs in SD; with the same specification, they achieve lower validation throughput than peers in HK and ML.
In OS the total throughput is much lower. The bottleneck, however, is not the bandwidth of the ordering service but the single-TCP connection bandwidth from OS to TK, as our netperf measurement suggests. Hence, the true benefits of gossip in OS cannot be observed; we attribute the slight improvement in throughput in OS in the experiment with gossip to fewer TCP connections running from OS to TK.

%% file: apps.tex
\section{Applications and Use Cases}
\label{sec:apps}

Major cloud operators already offer (or have announced) ``blockchain-as-a-service'' running \HLF, including Oracle, IBM, and Microsoft.
Moreover, \HLF currently powers more than 400 prototypes and proofs-of-concepts of distributed ledger technology and several production systems, across different industries and use cases~\cite{peterson2018}. Examples include a food-safety network~\cite{aitken2017}, cloud-service blockchain platforms for banking~\cite{fujitsu2017}, and a digital global shipping trade~\cite{groenfeldt2018} solution. In this section, we illustrate some real use cases where \HLF has been deployed.

\paragraph{Foreign exchange (FX) netting}
A system for bilateral payment netting of foreign exchange runs on \HLF. It uses a \HLF channel for each pair of involved client institutions for privacy.  A special institution (the ``settler'') responsible for netting and settlement is a member of all channels and runs the ordering service.  The blockchain helps to resolve trades that are not settling and maintains all the necessary information  in the ledger.  This data can be accessed in real time by clients and helps with liquidity, resolving disputes, reducing exposures, and minimizing credit risk~\cite{IBMCLS17}.

\paragraph{Enterprise asset management (EAM)}
This solution tracks hardware assets as they move from manufacturing to deployment and eventually to disposal, capturing additionally licenses of software assets associated with the hardware. The blockchain records the various life-cycle events of assets and the associated evidence. The ledger serves as a transparent system of record between all participants who are involved with the asset, which improves the data quality that traditional solutions struggle with. The multi-party consortium blockchain runs among the manufacturer, shippers, receivers, customers, and the installers. It uses a 3-tiered architecture, with a user interface connecting through the \HLF client to the peers. A detailed description of the first version is available online~\cite{IBMbeam17}.

\paragraph{Global cross-currency payments}
In collaboration with Stellar.org and KlickEx Group, IBM has operated a cross-currency payment solution since October 2017, which processes transactions among partners in the APFII organization in the Pacific region~\cite{IBMpayment17}.
The \HLF blockchain records financial payments in the form of transactions endorsed by the participants, together with the conditions they agree on.   All appropriate parties have access and insight into the clearing and settlement of financial transactions.

The solution is designed for all payment types and values, and allows financial institutions to choose the settlement network.  In particular, settlement may use different methods, and \HLF makes a decision on how to settle a payment, depending on the configuration of the participants.  One possible kind of settlement is through Lumens (Stellar's cryptocurrency), other ways are based on the type of the traded financial instrument.

%% file: relatedwork.tex
\section{Related Work}
\label{sec:relatedwork}

The architecture of \HLF resembles that of a middleware-replicated database as pioneered by Kemme and Alonso~\cite{KemmeA00}.  However, all existing work on this addressed only crash failures, not the setting of distributed trust that corresponds to a BFT system.  For instance, a replicated database with asymmetric update processing~\cite[Sec.~6.3]{DBLP:series/synthesis/2010Kemme} relies on one node to execute each transaction, which would not work on a blockchain. The execute-order-validate architecture of \HLF can be interpreted as a generalization of this work to the Byzantine model, with practical applications to distributed ledgers.

Byzantium~\cite{GarciaRP11} and HRDB~\cite{VandiverBLM07} are two further predecessors of \HLF from the viewpoint of BFT database replication.  Byzantium allows transactions to run in parallel and uses active replication, but totally orders \textsc{BEGIN} and \textsc{COMMIT/ROLLBACK} using a BFT middleware.  In its optimistic mode, every operation is coordinated by a single master replica; if the master is suspected to be Byzantine, all replicas execute the transaction operations for the master and it triggers a costly protocol to change the master.  HRDB relies in an even stronger way on a correct master.  In contrast to \HLF, both systems use active replication, cannot handle a flexible trust model, and rely on deterministic operations.

In \emph{Eve}~\cite{Kapritsos0QCAD12} a related architecture for BFT SMR has also been explored.  Its peers execute transactions concurrently and then verify that they all reach the same output state, using a consensus protocol.  If the states diverge, they roll back and execute operations sequentially. Eve contains the element of independent execution, which also exists in \HLF, but offers none of its other features.

A large number of distributed ledger platforms in the permissioned model have come out recently, which makes it impossible to compare to all (some prominent ones are Tendermint~\cite{TendermintURL}, Quorum~\cite{QuorumURL}, Chain Core~\cite{ChainURL}, Multichain~\cite{MultichainURL},
Hyperledger Sawtooth~\cite{SawtoothURL},
the Volt proposal~\cite{SettyBZRV17}, and more, see references in recent over\-views~\cite{DLZCOW17,CachinV17}).  All platforms follow the order-execute architecture, as discussed in Section~\ref{sec:background}.  As a representative example, take the Quorum platform~\cite{quorum17}, an enterprise-focused version of Ethereum.  With its consensus based on \emph{Raft}~\cite{Ongaro:2014:SUC:2643634.2643666}, it disseminates a transaction to all peers using gossip and the Raft leader (called \emph{minter}) assembles valid transactions to a block, and distributes this using Raft.  All peers execute the transaction in the order decided by the leader. 
Therefore it suffers from the limitations mentioned in Sections~\ref{sec:introduction}--\ref{sec:background}.

%% file: conclusion.tex
\section{Conclusion}
\label{sec:conclusion}

Fabric is a modular and extensible distributed operating system for running
permissioned blockchains.  It introduces a novel architecture that
separates transaction execution from consensus and enables policy-based
endorsement and that is reminiscent of middleware-replicated databases.

Through its modularity, Fabric is well-suited for many further
improvements and investigations.  Future work will address (1) performance
by exploring benchmarks and optimizations, (2) scalability to large
deployments, (3) consistency guarantees and more general data models, (4)
other resilience guarantees through different consensus protocols, (5)
privacy and confidentiality for transactions and ledger data through
cryptographic techniques, and much more.

\section*{Acknowledgments}

We thank Pramod Bhatotia and the anonymous reviewers for their very
insightful and constructive comments.

This work was supported in part by the European Union's Horizon 2020
Framework Programme under grant agreement number~643964 (SUPERCLOUD) and in
part by the Swiss State Secretariat for Education, Research and Innovation
(SERI) under contract number~15.0091.